\documentclass{aa}  
\usepackage{natbib}
\bibpunct{(}{)}{;}{a}{}{,}
\usepackage[varg]{txfonts}
\usepackage[colorlinks,allcolors=blue]{hyperref}
\hypersetup{
    colorlinks=true,
    linkcolor=blue,
    filecolor=magenta,      
    urlcolor=blue,
    citecolor=blue
}
\usepackage{ulem}
\usepackage{orcidlink}
\begin{document} 

\title{An efficient unsupervised classification model for galaxy morphology: Voting clustering based on coding from ConvNeXt large model}

   \author{
           Guanwen Fang\orcidlink{0000-0001-9694-2171}\inst{1}
           \and
           Yao Dai\orcidlink{0000-0002-4638-0235}\inst{1}
           \and
           Zesen Lin\orcidlink{0000-0001-8078-3428}\inst{2}
           \and
           Chichun Zhou\orcidlink{0000-0002-5133-2668}\inst{3}
           \and
           Jie Song\orcidlink{0000-0002-0846-7591}\inst{4,5}
           \and
           Yizhou Gu\orcidlink{0000-0003-3196-7938}\inst{6}
           \and
           Xiaotong Guo\orcidlink{0000-0002-2338-7709}\inst{1}
           \and
           Anqi Mao\orcidlink{0009-0002-3240-4408}\inst{3}
           \and
           Xu Kong\orcidlink{0000-0002-7660-2273}\inst{4,5}
           }

   \institute{Institute of Astronomy and Astrophysics, Anqing Normal University, Anqing 246133, People's Republic of China,
              \email{wen@mail.ustc.edu.cn}
        \and Department of Physics, The Chinese University of Hong Kong, Shatin, N.T., Hong Kong S.A.R., China 
        \and School of Engineering, Dali University, Dali 671003, People's Republic of China \email{zhouchichun@dali.edu.cn}
        \and Department of Astronomy, University of Science and Technology of China, Hefei 230026, China, \email{xkong@ustc.edu.cn}
        \and School of Astronomy and Space Science, University of Science and Technology of China, Hefei 230026, People's Republic of China
        \and Tsung-Dao Lee Institute, and Key Laboratory for Particle Physics, Astrophysics and Cosmology, Ministry of Education,  Shanghai Jiao Tong University, Shanghai 200240, China
              }

\date{Received ---;  accepted ---}

 \abstract
{By combining unsupervised and supervised machine learning methods, we have proposed a framework, called \texttt{USmorph}, to carry out automatic classifications of galaxy morphologies. In this work, we update the unsupervised machine learning (UML) step by proposing an algorithm based on ConvNeXt large model coding to improve the efficiency of unlabeled galaxy morphology classifications.
The method can be summarized into three key aspects as follows: (1) a convolutional autoencoder is used for image denoising and reconstruction and the rotational invariance of the model is improved by polar coordinate extension; (2)  utilizing a pre-trained convolutional neural network (CNN) named ConvNeXt for encoding the image data. The features were further compressed via a principal component analysis (PCA) dimensionality reduction; (3) adopting a bagging-based multi-model voting classification algorithm to enhance robustness. We applied this model to I-band images of a galaxy sample with $I_{\rm mag}< 25$ in the COSMOS field. 
Compared to the original unsupervised method, the number of clustering groups required by the new method is reduced from 100 to 20. Finally, we managed to classify about 53\% galaxies, significantly improving the classification efficiency. To verify the validity of the morphological classification, we selected massive galaxies with $M_{*}>10^{10}M_{\sun}$ for morphological parameter tests. The corresponding rules between the classification results and the physical properties of galaxies on multiple parameter surfaces are consistent with the existing evolution model. Our method has demonstrated the feasibility of using large model encoding to classify galaxy morphology, which not only improves the efficiency of galaxy morphology classification, but also saves time and manpower. Furthermore, in comparison to the original UML model, the enhanced classification performance is more evident in qualitative analysis and has successfully surpassed a greater number of parameter tests. The enhanced UML method will support the Chinese space station telescope in the future.}

\keywords{Galaxy: general -- Galaxy: structure -- galaxies: statistics -- Astrostatistics techniques -- Astronomy data analysis }
\titlerunning{An Efficient Unsupervised Classification Model}
\authorrunning{Fang et al.}
\maketitle

\section{Introduction}\label{sec:1}

In recent years, with the advancement of observational techniques and the development of observation instruments, several large-scale galaxy mapping projects such as the Sloan Digital Sky Survey (SDSS; \citealt{2002AJ....123..485S}) and the Cosmic Evolution Survey (COSMOS; \citealt{Scoville_2007}) have been initiated or completed, significantly increasing the number of observed galaxies (\citealt{2006ApJ...652..963R, Scoville_2007}). The exponential growth of data necessitates the development of more efficient data processing techniques to accomplish galaxy classification tasks.

The application of machine learning methods in classifying galaxy morphology has been increasing. 
\cite{10.1111/j.1365-2966.2010.16713.x} used artificial neural networks to classify galaxies labeled by Galaxy Zoo,  \cite{Dieleman_2015} introduces deep neural network models into galaxy morphology classification, achieving high-precision classification of Galaxy Zoo data. The study by
\cite{10.1093/mnras/stx2976} applied four statistical techniques (support vector machines, classification trees, random forest classification trees, and neural networks) to test the feasibility of automated algorithms for galaxy classification. \cite{article} proposed a residual-network variant of galaxy morphology classification to be applied to large-scale samples, \cite{Ghosh_2020} and \cite{10.1093/mnras/stab594}  both developed different supervised learning algorithms for the field of galaxy classification.

However, existing supervised machine learning algorithms often necessitate a substantial amount of precisely labeled data \citep{Schmarje_2021}, frequently sourced from visual categorization. Training the network typically demands prior knowledge of the labels, which could be time-consuming and computationally intensive, limiting the full exploitation of machine learning advantages \citep{Song_2024}. Furthermore, utilizing nearby galaxies as training samples to classify the morphology of high-redshift galaxies is only partially reasonable, as the morphological evolution of galaxies varies across different stages \citep{Tohill_2024}.
Therefore, we need to develop a new generation of efficient machine learning methods to mitigate labels' influence on galaxy classification results, such as the impact of uneven label categories and label labeling errors \citep{Tohill_2024}.

To achieve the above goal and improve the classification efficiency of machine learning, self-supervised learning \citep{9770283, DBLP:journals/corr/abs-2305-13689}, and unsupervised learning-based \citep{Schmarje_2021, DBLP:journals/corr/abs-2311-08995, DBLP:conf/iclr/DosovitskiyB0WZ21} approaches are proposed. Self-supervised learning models can extract features from intrinsic sample signals without requiring data labeling, thus reducing the dependence on manual labeling. Unsupervised machine learning methods do not require data labeling at all and focus on exploring the inherent structures and patterns within the unlabeled data through various techniques \citep{Hou2021AnUD}. This paper aims to introduce a rapid and effective unsupervised machine learning clustering algorithm, which is straightforward and user-friendly, to eliminate the influence of prior labeling on the model.

Some deep clustering frameworks \citep{Ajay2022UnsupervisedHM} can enhance feature extraction in the absence of data labels, such as SimCLR \citep{10.5555/3524938.3525087}, SWAV \citep{10.5555/3495724.3496555}, and SCAN \citep{10.1007/978-3-030-58607-2_16}. By combining the advantages of traditional convolutional neural networks (CNNs) with modern design improvements, ConvNeXt \citep{9879745} is also an efficient visual model that can proficiently extract features from similar fields. ConvNeXt represents an advance on the ResNet architecture \citep{7780459}, employing a training approach similar to Swin Transformer \citep{9710580}. It refines the main structure of ResNet by integrating a patchy layer and broadening the width of the network through convolutional grouping. This design combines the advantages of convolutional networks and Swin Transformer \citep{9710580}. By combining the robust feature extraction capability of CNN with the self-attention mechanism of Swin Transformer, long-term dependency relationships can be effectively simulated. ConvNeXt shows superior performance in image classification tasks \citep{9879745}. Compared to pure Transformer-based models such as Vision Transformer (ViT; \citealt{Paul}), ConvNeXt significantly reduces parameter number and computational overhead while maintaining superior performance. By pre-trained the ConvNeXt model with the ImageNet-22K dataset, \cite{DBLP:journals/corr/abs-2311-08995}  developed an effective method for classifying fungi. This demonstrates the effective transfer learning capability of the pre-trained ConvNeXt model and provides a new approach for directly using the pre-trained ConvNeXt model in astronomical image classification.

With long-term efforts, we have established a framework named USmorph for automatic classification of galaxy morphology.
\cite{Zhou_2022} developed the original bagging-based multi-model voting clustering method (UML), which enables machines to cluster galaxies into 100 highly similar groups. \cite{Fang_2023} introduced a supervised algorithm (SML) that integrates polar coordinate unfolding technology with GoogleNet, achieving high accuracy in galaxy classification. \cite{Dai_2023} were the first to combine UML and Googlenet models, demonstrating the effectiveness of this combined approach for classification. \cite{Song_2024} further expanded the application data by six times by combining UML and SML, validating the model's robustness and enhancing the clustering capability of the UML algorithm to 50 groups.

In this paper, we improve the original UML model (\citealt{Zhou_2022}) and introduce a technique to encode data using a pre-trained ConvNeXt large-scale model (\citealt{9879745}), thereby enhancing the efficiency of galaxy image feature extraction. The model used in this work is also pre-trained on the ImageNet-22K dataset. We have made the project code available at \url{https://github.com/ydai-astro/eUML}.
Our method leverages the convolutional
autoencoder (CAE; \citealt{10.1007/978-3-642-21735-7_7}) and adaptive polar-coordinate transformation techniques (APCT; \citealt{Fang_2023}) for data preprocessing to reduce image noise and align image features. The ConvNeXt large model is employed to encode the preprocessed samples for feature extraction, followed by a PCA dimensionality reduction on the encoded data for further feature extraction. The goal is to maximize the extraction of practical information from the data, eliminate redundant information, and achieve dimensional compression. This process enables us to reduce unnecessary resource consumption while preserving only the essential features. We employ a bagging-based multi-model voting clustering method (\citealt{Zhou_2022}) to classify the encoded data, ensuring accurate data classification. Compared to the original UML method, the enhanced UML method reduces the machine clustering groups from 100 to 20, thus conserving more workforce and resources. We finally cluster 20 groups into five categories of galaxy morphology by visual classification. The advantage of unsupervised clustering and visual classification is that it is straightforward and efficient to judge categories with similar characteristics, making the judgment challenging to be biased. We validated the effectiveness of our classification model by data visualization using t-SNE technology and examining the distribution of galaxy morphological parameters (\citealt{7b54165e73a3424b8820136bcf61ca89,doi:10.1137/18M1216134}).
The organizational structure of this paper is as follows. Section \ref{sec:2} describes the dataset and our sample selection. Section \ref{sec:3} details our improved self-supervised machine learning classification model, including data preprocessing techniques, ConvNeXt large model-based coding methods and the Bagging-based multi-model voting techniques. Section \ref{sec:4} mainly shows the clustering results in detail and verifies the validity of the classification results through the galaxy shape parameter test. Section \ref{sec:5} summarizes the main conclusions and offers prospects for future research.

\section{Data and sample selection} \label{sec:2}
\subsection{COSMOS field}
The COSMOS field was strategically designed to delve into the intricate relationships between various cosmic phenomena \citep{Scoville_2007}, all within the redshift range of $0.5 < z < 6$. This comprehensive survey stands as a remarkable exploration, covering an expansive 2 $\rm deg^2$ and spanning a diverse range of wavelengths from X-ray to radio. \cite{Koekemoer_2007} conducted a thorough processing of the images using the STSDAS Multidrizzle software package \citep{2003hstc.conf..337K}.  The processed images have a pixel scale of $0\farcs03$. 

This study is specifically focused on leveraging the high-resolution images from the Hubble Space Telescope's Advanced Camera for Surveys (HST/ACS) in the F814W filter. Encompassing an area of approximately 1.64 $\rm deg^2$ within the COSMOS field, these images mark the largest continuous field observed by the HST/ACS. All experimental data in this work are based on HST I-band data.

\subsection{COSMOS2020 catalog}
The "Farmer" COSMOS2020 catalog \citep{Weaver_2022} provides extensive photometric data across 35 bands, ranging from the ultraviolet to near-infrared (UV-NIR). This catalog was utilized to characterize various physical properties of galaxies, including photometric redshift and stellar mass, through spectral energy distribution (SED) fitting.

To determine the redshifts, two distinct codes were used: EAZY \citep{Brammer_2008} and LePhare \citep{refId0}. The findings suggest that these different estimation methods yield consistent results. To enhance reliability within the specific magnitude range under consideration, we adopted the LePhare-derived redshifts \citep{Weaver_2022}. This redshift estimation process utilized a library of 33 galaxy templates obtained from \cite{10.1046/j.1365-8711.2003.06897.x} and \cite{Ilbert_2009}.

\subsection{Sample selection}
\begin{figure}
        \includegraphics[width=1\columnwidth]{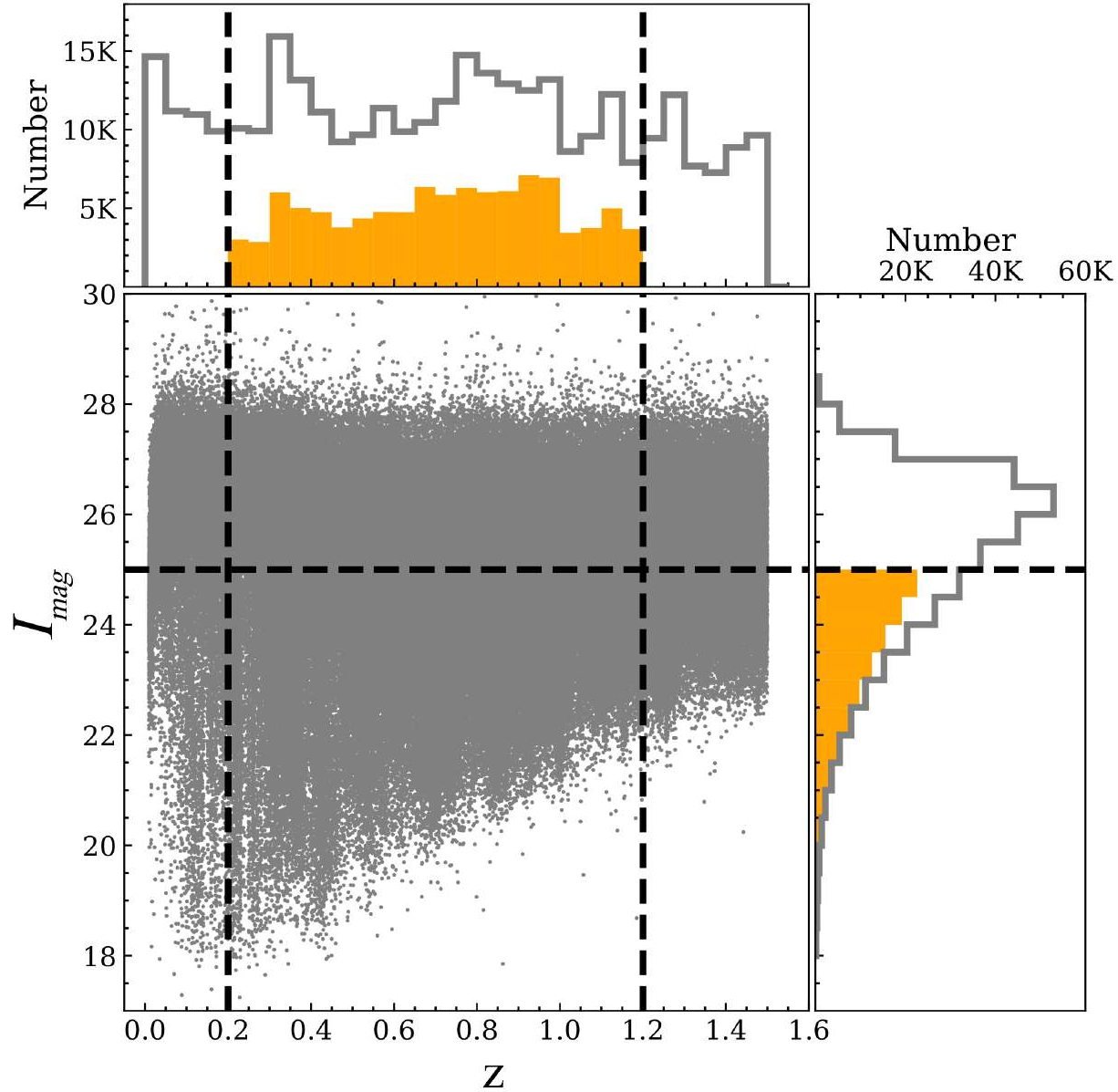}
 \centering
    \caption{Distribution of our sample in the I-band magnitude versus redshift diagram. The grey points represent the distribution of all galaxy samples in the COSMOS2020 Catalogue. In this work, we only consider samples with $I_{\rm mag} < 25$ and within the redshift range of $0.2 < z <1.2$. In the panels above and to the right of the image, the grey solid lines (yellow bars) show the distributions of $I_{\rm mag}$ and $z$ for all galaxy samples (the sample we selected) in the COSMOS2020 catalog.}
    \label{fig002}
\end{figure}

\begin{figure}
        \includegraphics[width=1\columnwidth]{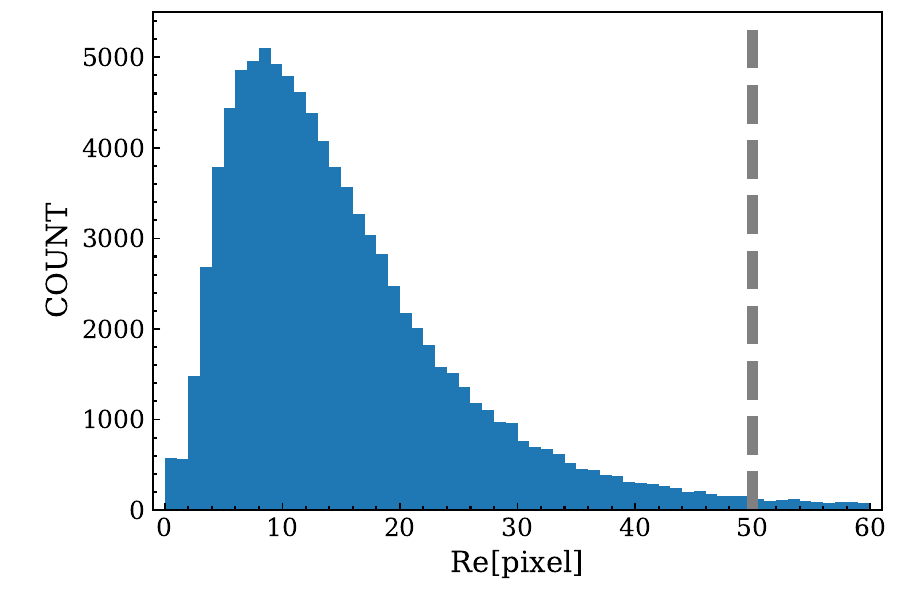}
 \centering
    \caption{Histogram of effective radius distribution of galaxies in the selected sample. The effective radius of 97.5\% of the samples is less than 50 pixels, it indicates that images with a size of 100 $\times$ 100 pixels could store the majority of information about galaxies, which means this cutout is suitable for this work.}

    \label{fig:1}
\end{figure}

In this study, our galaxy selection process from the COSMOS2020 catalog adhere to the following criteria: (1) galaxies with $I_{mag} < 25$ were chosen, excluding those that are too faint for accurate morphological measurements; (2) the redshift range of $0.2 < z < 1.2$ was applied to ensure morphological assessments were conducted in the rest-frame optical band; (3) the selection was limited to galaxies with $\rm FLAG_{COMBINE}=0$, ensuring reliability in photometric redshift and stellar mass estimations; and 4) additionally, galaxies with bad pixels and signal-to-noise ratios less than 5 (S/N<5) were excluded from the analysis. The resulting sample comprises a total of 99,806 galaxies. The redshift and $I_{\rm mag}$ distributions of our sample are shown in Fig. \ref{fig002}. The size of the captured image is 100 $\times$ 100 pixels. We measured the effective radius of galaxies using GALAPAGOS software \citep{Barden2012GALAPAGOSFP, 2022A&A...664A..92H}, which corresponds to the major-axis of an ellipse. Fig. \ref{fig:1} illustrates the distribution of effective radii across all our samples. Remarkably, 97.5\% of the samples exhibit an effective radius of less than 50 pixels, which ensures that our cutouts contain enough information about the galaxies to perform morphology classification.

\section{Enhanced UML method based on large pre-trained ConvNeXt model} \label{sec:3}
\begin{figure*}
        \includegraphics[width=2\columnwidth]{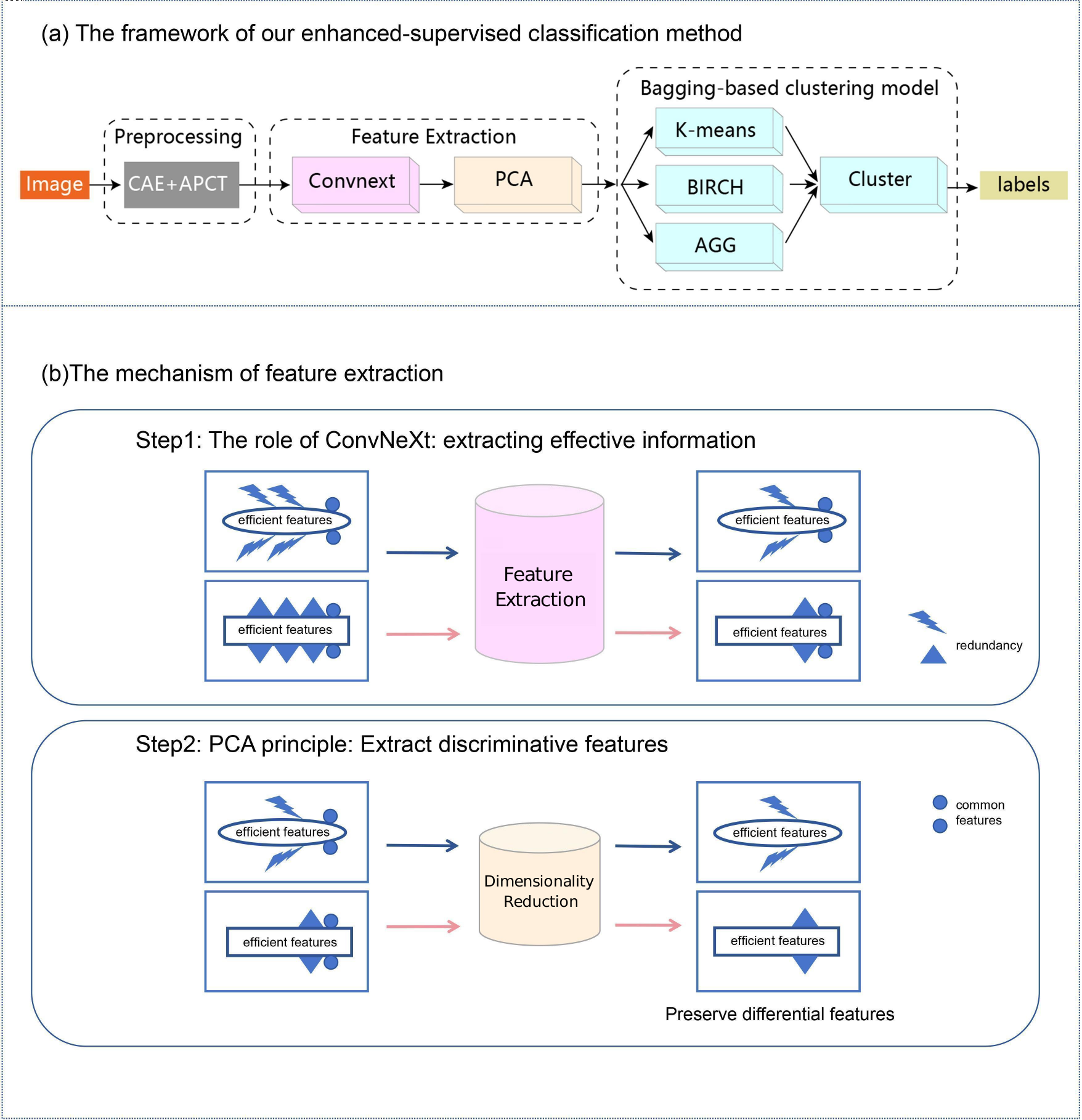}
        \centering
        \caption{Our framework for enhancing unsupervised classification and schematic diagram for feature extraction. Panel (a) represents the improved model flowchart, while Panel (b) represents the schematic of feature extraction. In the feature extraction process, we use the ConNeXt large model for data encoding to extract compelling features from the data. Subsequently, the next step involves PCA dimensionality reduction to eliminate redundant information in the image. This combination effectively extracts features while retaining only basic information, optimizing computing resources.}
    \label{fig:2}
\end{figure*}

In this section we delve into the intricate architecture of the enhanced UML technique. Fig. \ref{fig:2} illustrates the specific process of the enhanced UML method and the main principles of feature extraction. The initial image undergoes preprocessing before being fed into the feature extraction module, where it is subjected to feature extraction and dimensionality reduction processes. Subsequently, a multi-model voting clustering approach is employed. Here, we provide an exhaustive overview.

\subsection{Data preprocessing}
The CAE mitigates noise by recreating the input images, ensuring the fundamental information of the image is retained while extracting salient features to diminish data dimensions. Given these attributes, we employ CAE as a data preprocessing tool, wherein its competencies in reducing image noise and compressing dimensions are manifested throughout a sequence of preparatory procedures for our work \citep{Zhou_2022, Fang_2023, Dai_2023, Song_2024}. In addition, by combining autoencoders and clustering algorithms, the recognition of gravitational lenses \citep{10.1093/mnras/staa1015}, and effective morphological classification \citep{10.1093/mnras/stab734} have also demonstrated the practicality and superiority of CAE in multiple research topics. 

\begin{figure*}
\includegraphics[width=2\columnwidth]{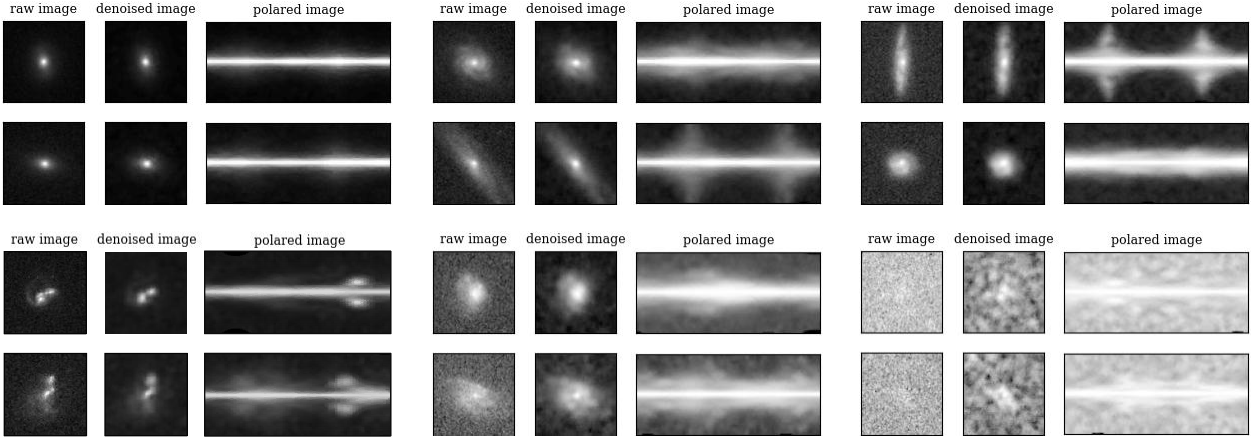}
    \caption{Some examples of image preprocessing, including six sets of images showing the comparison between preprocessed images of different shapes of galaxies and the original image. The original image is on the left side of each group of images, in the middle is the image after CAE denoising, and on the right is the image after the polar coordinate unfolding. The data after CAE denoising effectively preserves the characteristics of galaxies while eliminating artifacts such as noise and artifacts in the image. The subsequent APCT processing, after polar coordinate unfolding, inverts the image up and down and concatenates it, enhancing the brightest area in the centre and highlighting the significant differences between different categories of galaxies, thereby enhancing the rotational invariance of the model. From the graph, we can see that galaxies with different morphologies exhibit significant differences after preprocessing.    
    }
    \centering
    \label{fig:3}
\end{figure*}

In Fig. \ref{fig:3}, the left image of each group represents the original image, while the middle image represents the reconstructed image after CAE denoising. The figure shows significantly reduced background noise and preserves the structural characteristics of the galaxy's interior. Table \ref{tab1} shows the convolution kernel sizes and structure of the CAE used in this work. The decoding process of the CAE is exactly the reverse of the encoding process. The model performs best when the convolution kernel size in experiments is set to be 5 $\times$ 5.

The classification of galaxy morphology should be rotational invariant. To enhance the rotation invariance of the model, traditional data augmentation methods reduce the impact of image angles through simple rotation, sorting, and so on. An image needs to be rotated into multiple images with different angles for data augmentation, which significantly increases the computational burden and cannot eliminate the impact of rotation. The APCT technique developed by \cite{Fang_2023} can effectively enhance the rotation invariance of the model, while saving computational power. This technique unfolds the image from the centre's brightest to the darkest point by adding pixels every 0.05 degrees counterclockwise and mirrors and concatenates the unfolded image. The operation not only enhances the most prominent features in the centre, but also eliminates the angle information in the image, focusing more on the differential information of the image. Compared with traditional data augmentation, it effectively saves computational resources, enhances model rotation invariance and is more suitable for large-scale data surveys. 
Using APCT technology to process images could highlight the feature differences, which is of benefit to label alignment and thus the subsequent classification.

In Fig. \ref{fig:3}, the right column of each image group shows the images processed by APCT and there are significant differences in the morphology of galaxies after APCT processing while retaining distinct features. Galaxies dominated by nuclear spheres have a clear, narrow white channel in the middle of the image after unfolding in polar coordinates, without any other structures. After unfolding, galaxies with both a nuclear sphere and a disk structure have white channels and additional edge information due to different disk shapes. The brighter the nuclear sphere, the more pronounced the white channels are; vice versa, the more pronounced the edge information is. Galaxies with multi-core structures exhibit significant differences from other galaxies after unfolding, except for the white channel. These similar and different features provide a foundation for subsequent model classification.

Overall, the combination of CAE and APCT effectively reduces background noise, extracts compelling features, and eliminates angle information in the images, improving the model's classification efficiency and rotation invariance.

\begin{table}
\caption{CAE architecture used in this work}             
\label{tab1}      
\centering          
\begin{tabular}{c c c c }   
\hline\hline                     
 Operation & Dimension & Filter Size  & Stride \\
\hline                    
 Input       & $100\times 100 \times 1$ & ... & ... \\
 Convolution & $100\times 100 \times 8$ & $5\times 5$ & ... \\
 Maxpooling  & $50 \times 50  \times 8$ & $2\times 2$ & $2\times 2$ \\
 Convolution & $50 \times 50  \times 8$ & $5\times 5$ & ... \\
 Maxpooling  & $25 \times 25  \times 8$ & $2\times 2$ & $2\times 2$ \\
 Unfolding   & 10000   & ... & ... \\
 Full connection & 40 & ... & ... \\
\hline                  
\end{tabular}
\end{table}

\subsection{Feature extraction and dimensionality reduction}

This section primarily elucidates the principles and parameters pertinent to model feature extraction and dimensionality reduction, as delineated below.
ConvNeXt was introduced by \cite {9879745}, representing an evolution beyond ResNet convolutional networks \citep{7780459}. It combines training techniques similar to the Swin Transformer, redefining the ResNet architecture by introducing patch layers and expanding network width through convolutional grouping.

\begin{figure}\includegraphics[scale=0.3]{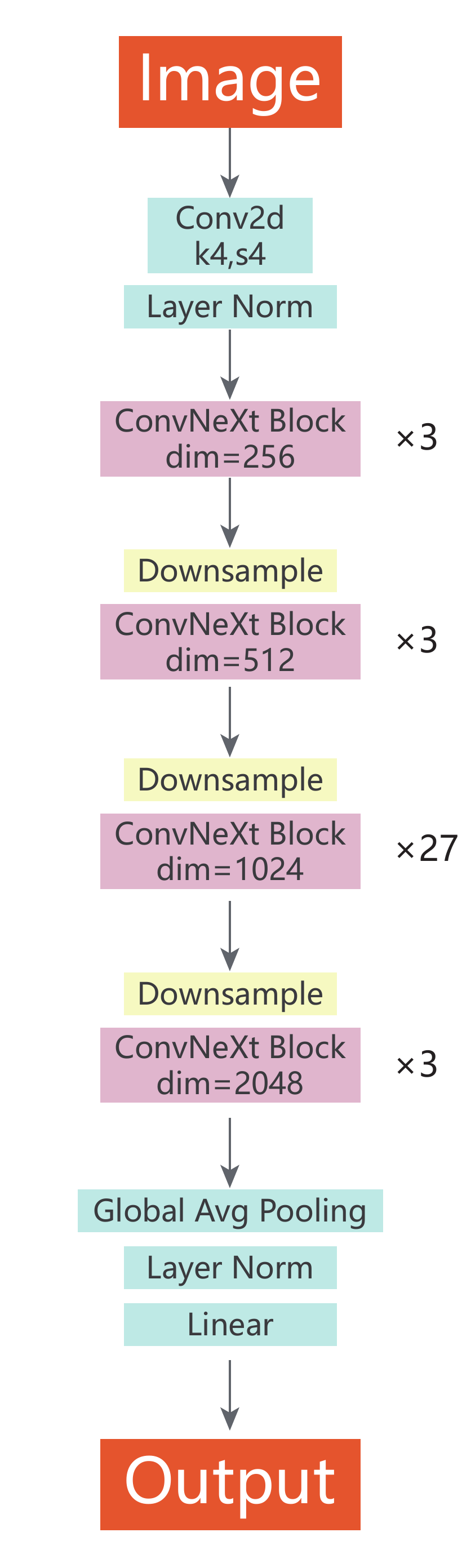}
\centering
    \caption{Framework of ConvNeXt model. ConvNeXt adopts a modular design approach similar to Transformer, dividing the network structure into multiple identical modules for easy expansion and maintenance. The network adopts a standardized connection method between each module, which facilitates information transmission and gradient flow. Although ConvNeXt does not introduce particularly complex or novel structures, it achieved good performance by effectively integrating and applying existing methods.}
    \label{fig:4}
\end{figure}

This work uses a pre-trained ConvNeXt model of which the specific structure is shown in Fig. \ref{fig:4}. The numbers of channels corresponding to each layer of blocks are 256, 512, 1024, and 2048, respectively, while the numbers of stacked blocks in each layer are 3, 3, 27, and 3, respectively.
The convolution kernel sizes of the block and downsampling layers are fixed throughout the ConvNeXt architecture. However, given the increasing dimensions of the block layers during the learning (i.e., from 256 to 2048), the convolution kernel size within each block layer is 7$\times$7 with a step size of 1, and the convolution kernel size of each downsampling layer is 2$\times$2 with a step size of 2.
The activation function employed is GELU, which amalgamates the benefits of both convolutional networks and Transformers.

This variation in stride facilitates the network in progressively extracting higher-level features during the learning phase, thereby enhancing the model's learning efficiency.  This not only substantially diminishes the parameters and computational burden but also sustains the robust feature extraction prowess of convolutional neural networks. Upon encoding with the ConvNeXt model, the image is transformed into a 2048-dimensional vector, effectively minimizing redundancies and extracting salient features from the image. To further enhance the efficiency of feature extraction, we aim to retain the differential characteristics between distinct images, while eliminating certain standard features; thereby achieving additional dimensionality reduction. Therefore, we used PCA to further reduce the dimension of feature vectors to achieve efficient feature extraction and save computing resources. 

Principal component analysis (PCA; \citealt{MACKIEWICZ1993303}) is a dimensionality reduction and feature extraction method. The method maps the original data to a new coordinate system via a linear transformation to find the data's main structures and patterns, during which dimensionality reduction is achieved by identifying the principal components of the data and concentrating the variance on a few principal components.
The primary objective of using PCA in learning is to minimize the dimensionality of data while maximizing the retention of original data information.
PCA accomplishes dimensionality reduction by computing the covariance matrix of the data, along with its eigenvalues and eigenvectors. The eigenvalues signify the variance of the data in various directions, whereas the eigenvectors denote these directions. By selecting the eigenvectors corresponding to the largest k eigenvalues, the original data can be projected onto a new k-dimensional space. By choosing different k values, a balance can be struck between reducing dimensionality and preserving information. Typically, these selected k eigenvalues account for the majority of the variance in the data, thereby preserving the main information of the data. We have calculated the eigenvalues and found that reducing the dimensionality from 2048 to 1500 can preserve the most effective information. Therefore, the final feature vector is extracted to 1500 dimensions,  retaining ample effective information.

As shown in Panel (b) of Fig. \ref{fig:2}, we improve the efficiency of model feature extraction by using a two-step strategy.
Firstly, we extract compelling features from the image to reduce redundant features. After the convolutional network structure of ConvNeXt, the image features are effectively extracted, and the redundant and repeated features within a single image are removed. After the secondary feature extraction of PCA dimensionality reduction, the same features between different images are removed and the difference features between images are retained (panel b in Fig. \ref{fig:2}).

\subsection{Clustering process}

The bagging-based multi-model voting classification method has been repeatedly proven to have good robustness and provide reliable classification results in previous studies \citep{Zhou_2022, Dai_2023, Song_2024}, at the cost of discarding some controversial samples.\footnote{ Note that the discarded galaxies in this step could be re-classify in the SML step of our USmorph framework (see \citealt{Fang_2023} for more details), which means that none of the galaxies would be discarded finally in our classification scheme.} Three independent clustering models were used to process data: a hierarchical balanced iterative reduction and clustering algorithm \citep{10.1145/233269.233324, Peng2020}, k-means clustering algorithm \citep{8ddb7f85-9a8c-3829-b04e-0476a67eb0fd}, and hierarchical aggregation clustering algorithm \citep{10.1093/comjnl/26.4.354, Murtagh_2014}. Each model independently classifies samples into 20 groups. The voting clustering algorithm aligns labels by setting the ones generated from the k-means algorithm as the fiducial labels. Once the categories are aligned, a voting classification decision is made. The corresponding groups will be retained only if at least two of the three models reach a consensus, The schematic diagram can be seen in Fig. 4 of \cite{Zhou_2022}. If no consensus is reached, the group classification is considered unreliable and excluded. Through this mixed clustering voting mechanism, 53216 galaxies were successfully classified into 20 groups.

During the voting-based clustering process, we experimented with clustering into 5, 10, and 20 groups, respectively.
After implementing the feature extraction scheme combining ConvNeXt large model encoding with PCA dimensionality reduction. the clustering algorithm can distinguish ellipsoids from other morphological types when voting for 5 or 10 groups. However, with that small group number, the algorithm failed to separate other galaxy morphologies.
When voting for 20 groups, we were able to group images with similar features, demonstrating effective discrimination. Consequently, we settled on 20 groups as the optimal group number. For comparison, the group number adopted in \cite{Zhou_2022} with the original UML method is 100. By utilizing the new feature extraction algorithm combined with PCA, only 20 groups are now required to achieve satisfactory classification results.

\begin{figure*} \includegraphics[width=2\columnwidth]{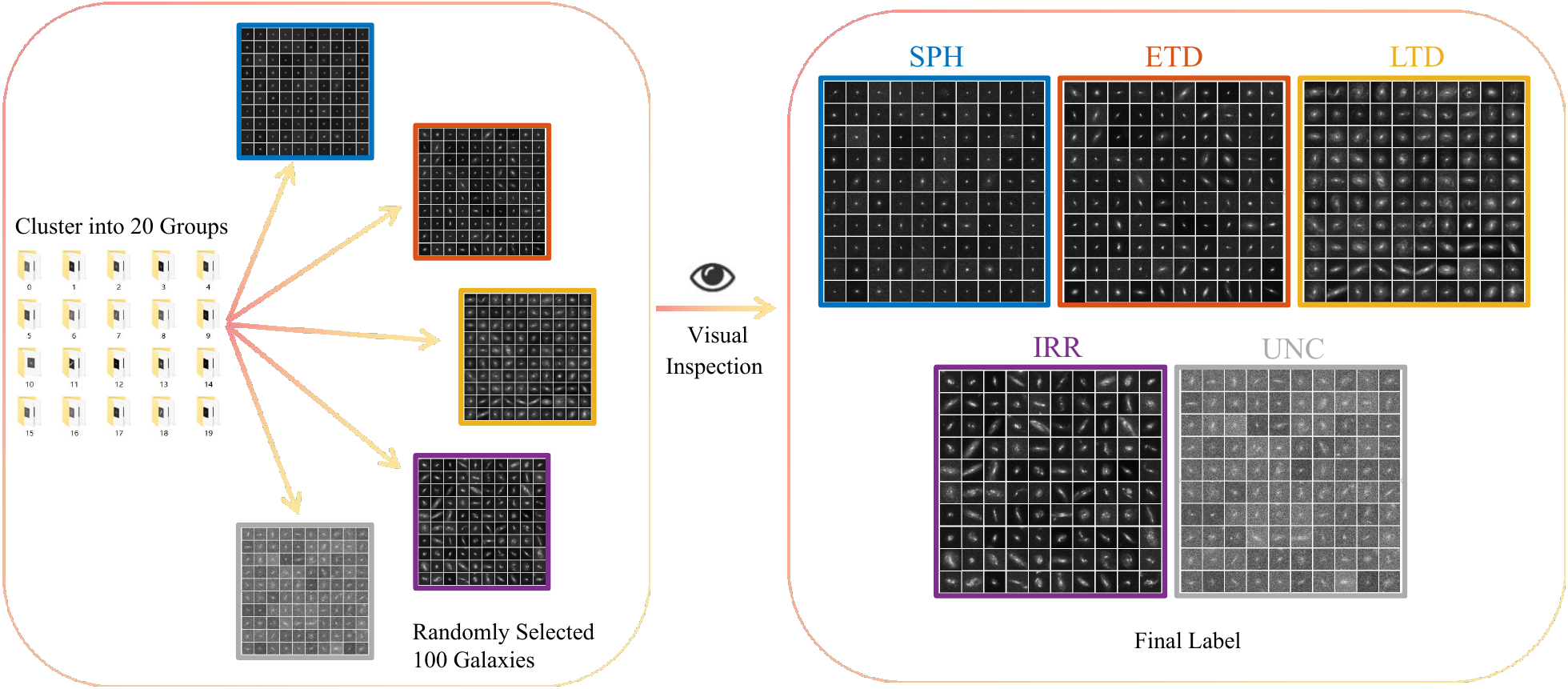}
    \caption{Visual categorization process diagram. From 20 machine clustering groups, we randomly selected 100 images from each group and placed them on a single canvas. Three experts voted on the classification, ultimately categorizing them into five physically meaningful categories. Within the randomly selected images, there was high similarity within each group and significant differences between different groups. This process demonstrated the high efficiency and convenience of our machine learning method, significantly saving time and effort.}
    \label{fig:5}
\end{figure*}

To further sort the obtained 20 groups into five distinct morphological categories that possess physical significance for future scientific exploration, visual classification is carried out.
We randomly select 100 galaxy images from each group and place them on the same panel. Three experts participated in this process, and when two or more experts made consistent judgments on the classification of the group of galaxies, we successfully classified them. Otherwise, it would have been deemed unclassifiable. Finally, the 20 groups of galaxies were classified into five categories with physical significance. Fig. \ref{fig:5} shows this classification process.

\begin{figure*} \includegraphics[width=2\columnwidth]{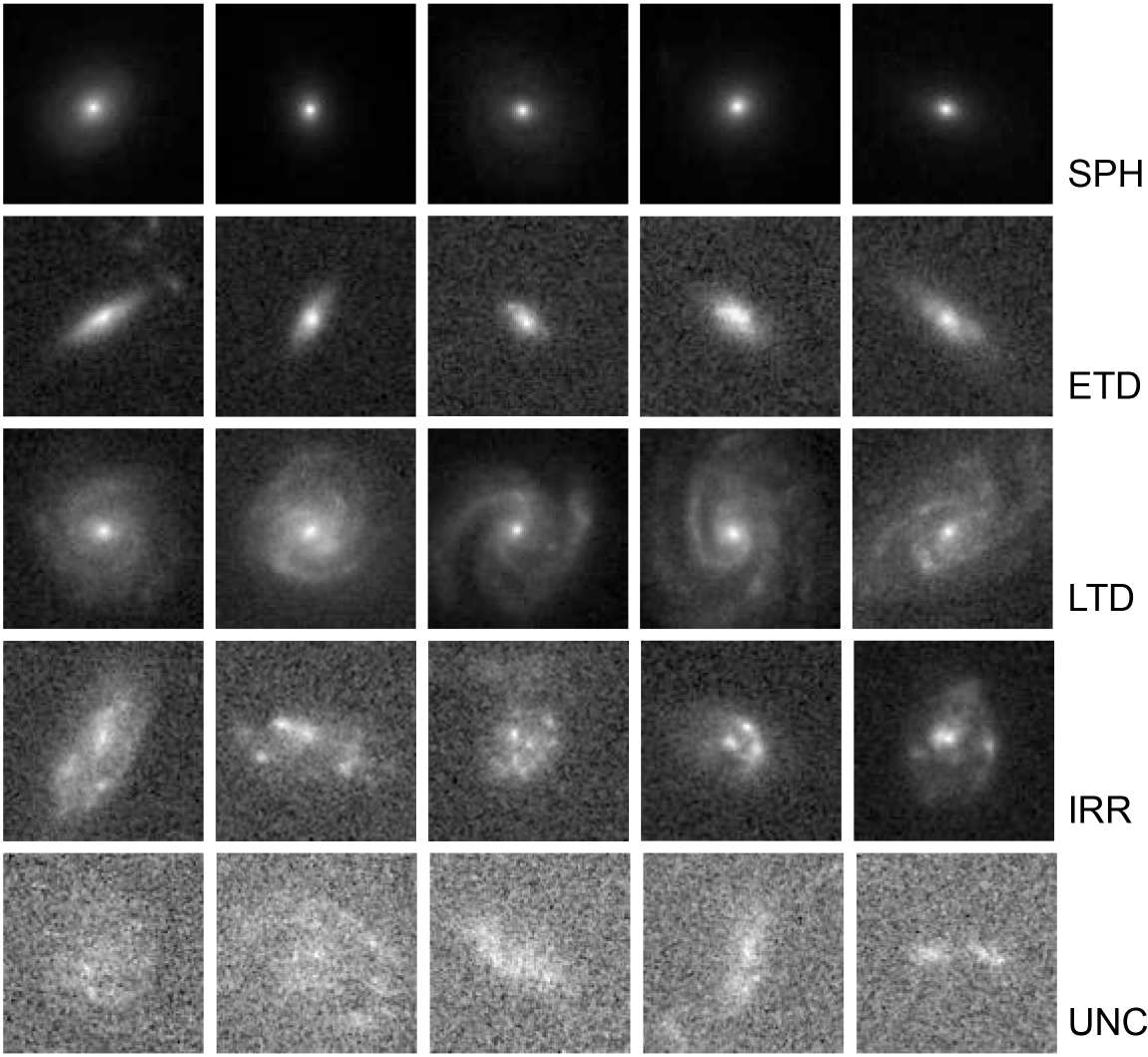}
\centering
    \caption{ Randomly selected examples of the five morphological categories.
    The images vividly illustrate the pronounced morphological distinctions between each galaxy type. SPH galaxies exhibit a unique spherical form with notably centralized brightness, whereas ETD galaxies feature a prominent nuclear sphere accompanied by a smaller disk-like shape. LTD galaxies are characterized by their well-defined spiral arms and a small central sphere. IRR galaxies display an irregular morphology, and UNC galaxies have an extremely low signal-to-noise ratio, which complicates the discernment of encapsulated information.}
    \label{fig:6}
\end{figure*}

These five morphological categories are spherical (SPH), early-type disk (ETD), late-type disk (LTD), irregular (IRR), and unclassified (UNC). SPHs have obvious spherical structures. ETDs have obvious nuclear spheres and smaller, regular disk structures, while LTDs have large and obvious disk structures attached, with visible spiral arms. IRRs do not have obvious regular shapes, and their nuclei are not obvious or have multiple nuclear structures. UNCs usually have low signal-to-noise ratios and cannot distinguish whether they contain galaxies or morphological features.
       
\section{Result and discussion} \label{sec:4}
In this section we present the final results of galaxy classification and validate its effectiveness through some tests.

\subsection{Classification result}

We finally successfully classified 53,612 (54\%) galaxies, including 10049 SPHs, 11763 ETDs, 8364 LTDs, 14433 IRRs, and 9103 UNCs. Randomly selected examples of the final classification results are exhibited in Fig. \ref{fig:6}, from which we can see that there is a clear differentiation in the morphologies of different types of galaxies.

\begin{figure*}
        \includegraphics[width=1.5\columnwidth]{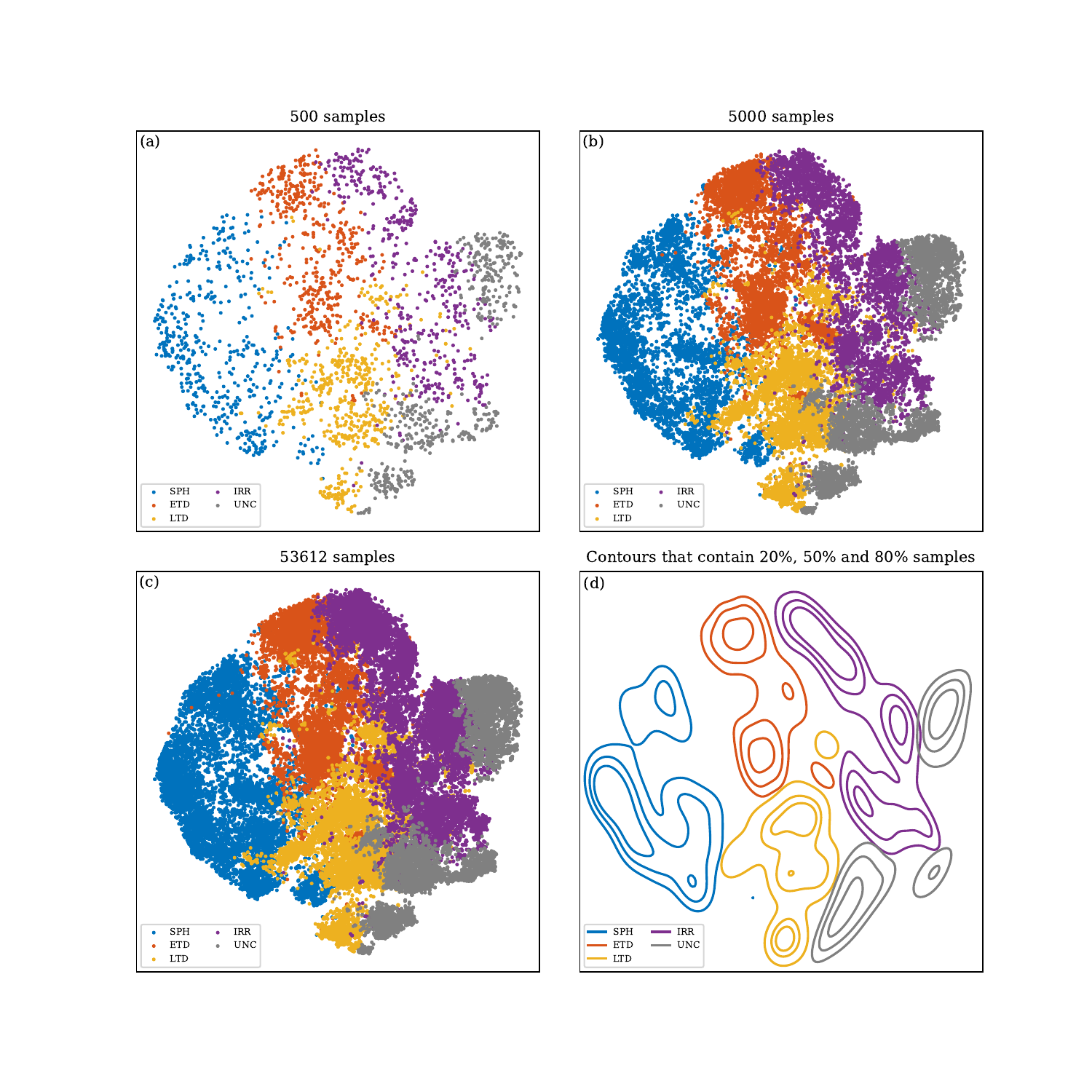}
 \centering
    \caption{t-SNE display of the final classification result. We use the t-SNE dimensionality reduction technique to perform two-dimensional mapping on the final classified 5 types of galaxies. Panels (a), (b), and (c) respectively display the dimensionality reduction results of randomly selected 500, 5000, and all samples. Panel (d) represents the contours that contain 20\%, 50\%, and 80\% of the well-classified samples, respectively.}
    \label{fig:7}
\end{figure*}

We used t-distribution random neighborhood embedding (t-SNE; \citealt{7b54165e73a3424b8820136bcf61ca89}) to illustrate the clustering results, which is a high-dimensional data visualization and dimensionality reduction technique. In Fig. \ref{fig:7}, we show the t-SNE distribution of the final clustering into five categories, randomly mapping 500, 5000 samples, and a total of 53612 fully classified data for two-dimensional mapping. The concentrated behavior of galaxies belonging to the same morphological types suggests that both machine clustering and visual classification show good classification performance. The same type of galaxy is distributed in adjacent spaces, while different galaxy mappings have different distribution trends in two-dimensional space. Sampling galaxies at different times ensures no overlapping effect between galaxies. In Panel (d) of Fig. \ref{fig:7}, the contour map represents the trend of galaxy clustering. Among various types of galaxies, 80\% of samples do not overlap after t-SNE, indicating the effectiveness of our clustering. Since the t-SNE test can only qualitatively test the distribution of classifications, we will now examine the effectiveness of classification from the perspective of physical properties.

\subsection{Test of morphological parameters} \label{subsec:morph}

Galaxies with different morphologies often possess distinct physical properties, which are intimately linked to the formation and evolution of galaxies (e.g., \citealt{10.1111/j.1365-2966.2007.12627.x, Gu_2018}). The distributions of galaxy morphology or structure parameters can be used to test the effectiveness of classification results (e.g., \citealt{Zhou_2022, Dai_2023, Yao_2023}). In this section we measure different morphology parameters to test the effectiveness of classification. Due to the unclear evolution mechanism of less massive galaxies, we only selected massive galaxies with $M_{*}>10^{10}~M_{\sun}$ for testing. UNC, which has difficult-to-measure morphological parameters, will not be considered for testing.

\subsubsection{Parametric measurements}

To obtain the morphological parameters of galaxies, we utilized the GALFIT package \citep{Peng_2002} and GALAPAGOS software \citep{Barden2012GALAPAGOSFP, 2022A&A...664A..92H} to fit the galaxy surface brightness profile with a single S\'{e}rsic model. This process allows us to measure the S\'{e}rsic index (n) and effective radius ($r_e$) for each galaxy.

\begin{figure*}
    \includegraphics[width=1.5\columnwidth]{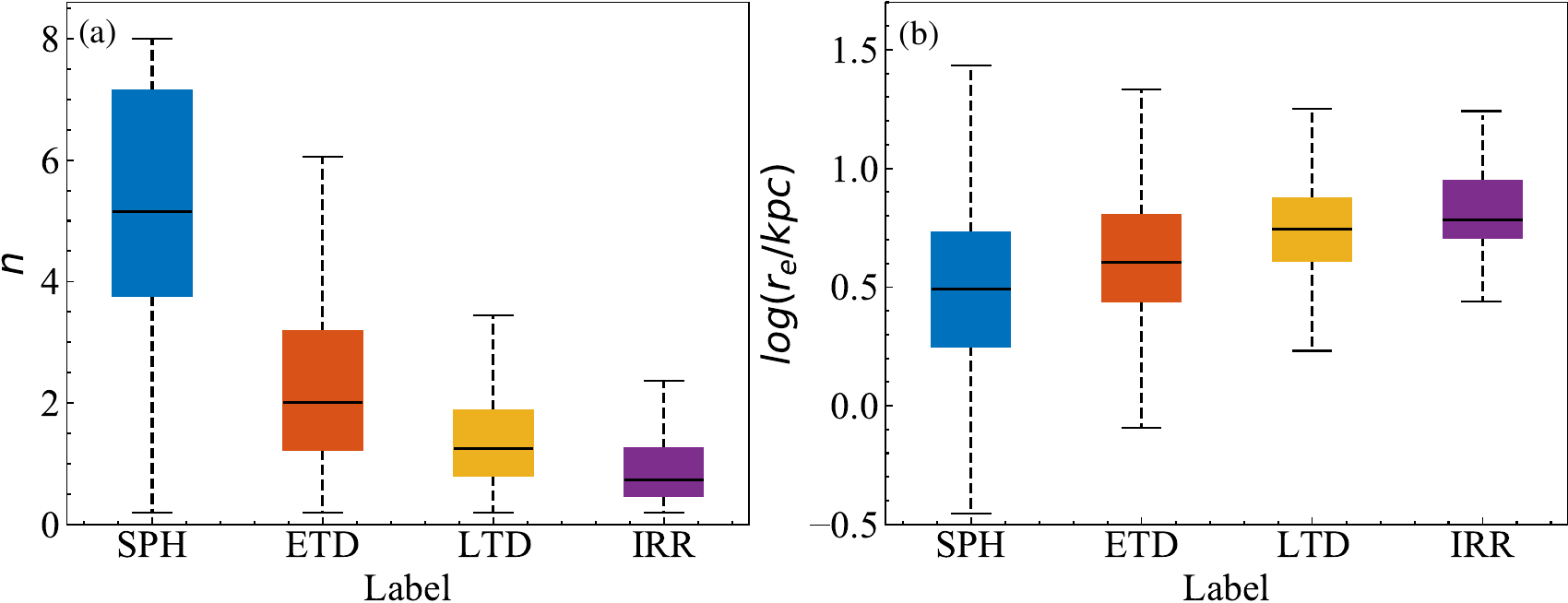}
    \centering
    \caption{Box plots of S\'{e}rsic index (left) and effective radii (right) for different morphological types of massive galaxies. The bar chart represents the values of S\'{e}rsic index and effective radii for different galaxy types. It can be seen that from SPHs to IRRs, the galaxy S\'{e}rsic index gradually decreases, and effective radii gradually increase, which further indicates the reliability of our classification results. The end of each box represents the 40\% upper and lower quartiles respectively, the ends of the dotted lines represent the maximum and minimum values, and the black horizontal line in the middle of the box represents the median.}
    \label{fig:8}
\end{figure*}

In Fig. \ref{fig:8}, we provide the distributions of the S\'{e}rsic index effective radius among the 53,612 galaxies that are successfully clustered using the improved UML method. The median S\'{e}rsic indices for SPHs, ETDs, LTDs, and IRRs are 4.4, 1.3, 1.1, and 0.8, respectively, demonstrating a gradual decrease. 
For $r_e$, the median values are 2.1, 2.8, 4.3, and 4.4kpc for SPH, ETD, LTD, and IRR galaxies, respectively, reflecting a gradual increase. Generally, the distribution of these two parameters aligns with the expected relationship with galaxy morphology \citep{2023ApJ...946L..15K}. This allows for differentiating different galaxy types in parameter space, validating the effectiveness of our clustering approach.

\subsubsection{Nonparametric measurements}

Using the statmorph\_csst \citep{Yao_2023}, we calculate the Gini coefficient (G)
and the normalized second moment of the brightest 20\% of the galactic flux ($M_{20}$) for all galaxies in the sample, as well as the CAS statistics and MID parameters \citep{Rodriguez_Gomez_2018}.

The G represents the degree of uniform distribution of light in the galaxy (e.g., \citealt{Lotz_2004, Lotz_2008}). It can be expressed as:
\begin{equation}
    G=\frac{1}{\overline{f}n_{pix}(n_{pix}-1)}\sum_{i=0}^n(2i-n_{pix}-1)f_i.
\end{equation}
In the context provided, $n_{pix}$ is the number of pixels comprising the galaxy,  while $f_i$ denotes the pixel flux values arranged in ascending order and $\overline{f}$ stands for the mean flux value across all pixels. 

$M_{20}$ represents the second moment of the brightest region of a galaxy \citep{Lotz_2004}, it can be expressed as:
\begin {equation}
M_{tot}=\sum _{i}^{n}M_{i}=\sum _{i}^{n}f_{i}[(x_{i}-x_{c})^{2}+(y_{i}-y_{c})^{2}]\
,\end{equation}

\begin {equation}
M_ {20} = \log_{10}\frac{\sum_i M_i} {M_{\rm tot}}, \ 
\mathrm{while}\ 
\sum_if_i < 0.2 f_{\rm tot},
\end{equation}
where ${M_{\rm tot}}$ represents the total second-order central moment, ${M_{\rm i}}$ represents the pixel flux values, $f_{\rm tot}$ represents the total flux of the galaxy, $f_i$ denotes the flux value of each pixel $i$, and $(x_i,\ y_i)$ signifies the position of pixel $i$, with $(x_c,\ y_c)$ indicating the center of the image \citep{Lotz_2004, Lotz_2008}.

The concentration (C) quantifies the distribution of light within a galaxy's central region relative to its outer regions (e.g., \citealt{Conselice_2000, Conselice_2003}). It can be expressed as:
\begin {equation}
C = 5 \log_{10} \left( \frac{r_{80}}{r_{20}} \right),
\end{equation}
where $r_{20}$ and $r_{80}$ represent the radii of circular apertures containing 20\% and 80\% of the galaxy’s light, respectively \citep{Conselice_2003}.

\cite{10.1093/mnrasl/sly054} introduced the $G_{2}$, which aims to identify and differentiate between elliptical and spiral galaxies by analyzing the symmetry of the flux gradient. It can be expressed as:
\begin{equation}
G_2 = \frac{V_A}{V}\times \left(2-\frac{|\Sigma_{k=1}^{V_A} v_k|}{\Sigma_{k=1}^{V_A}|v_k|}\right).
\end{equation}
Here, $V$ denotes the total number of local gradient vectors and $V_A$ represents the count of asymmetric vectors. The magnitude of the sum of all asymmetric vectors is given by $|\sum_{k=1}^{V_A} v_k|$, whereas the aggregate of the magnitudes of these asymmetric vectors is expressed as $\sum_{k=1}^{V_A}|v_k|$.

At the same time, we have also used some new measurement parameters. The multimode statistic (M) quantifies the ratio between the galaxy's two most prominent mass areas \citep{10.1093/mnras/stt1016, Rodriguez_Gomez_2018}. It can be expressed as:
\begin {equation}
M = \max_q \left( \frac{A_{q, 2}}{A_{q, 1}} \right),
\end{equation}
where $q$ is a value between 0 and 1 that represents the threshold for flux values in an image analysis. These represent the areas of the n groups of contiguous pixels sorted in descending order: $A_{q, 1}$ is the area of the largest group, $A_{q, 2}$ is the area of the second-largest group, and so on, until $A_{q, n}$ which is the area of the smallest group among the identified contiguous pixel clusters \citep{10.1093/mnras/stw252}.

The intensity statistic (I) indicates the ratio between the light of the galaxy's two brightest subdomains \citep{10.1093/mnras/stt1016, Rodriguez_Gomez_2018}. It can be expressed as:
\begin {equation}
I = \frac{I_{2}}{I_{1}}.
\end{equation}
The deviation statistic (D) signifies deviation, representing the distance between the weighted center of the image and the brightest peak \citep{10.1093/mnras/stt1016, Rodriguez_Gomez_2018}. It can be expressed as:
\begin {equation}
D = \sqrt{\frac{\pi}{n_{\rm seg}}[(x_{c}-x_{l})^{2}+(y_{c}-y_{l})^{2}]},
\end{equation}
where $(x_c, y_c)$ is the centroid of the image calculated using the pixels marked by the MID segmentation map, $(x_{l}, y_{l})$ is the brightest peak detected during the calculation of the $I$ statistic, while $n_{\text{seg}}$ is the count of pixels in the segmentation map, which is used to approximate the radius of the galaxy \citep{10.1093/mnras/stt1016}.

\begin{figure*}
        \includegraphics[width=1.5\columnwidth]{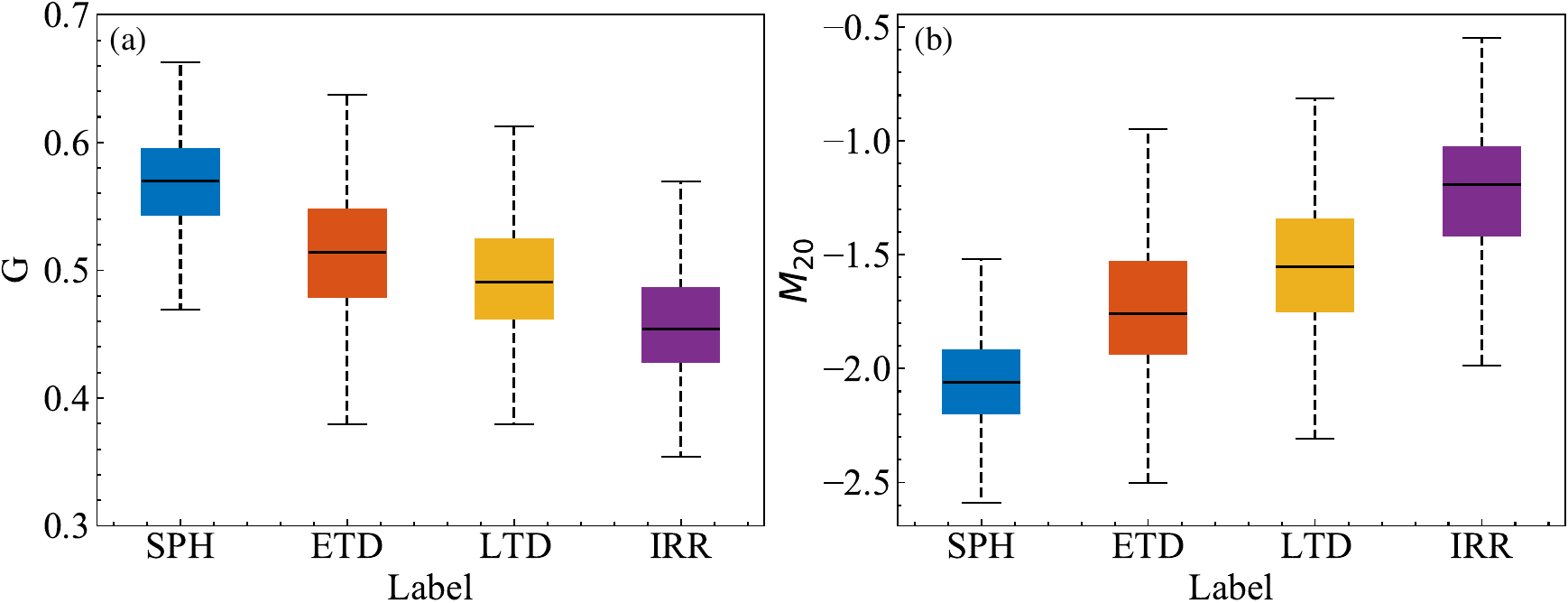}
 \centering
    \caption{Similar to Fig. \ref{fig:8} but for $G$ (left) and $M_{20}$ (right).
    It can be seen that from SPHs to IRRs, G gradually decreases, and $M_{20}$ gradually increases. It is consistent with the morphological evolution of galaxies, which further indicates the reliability of our classification results.}
    \label{fig:9}
\end{figure*}

Through the analysis of these parameters, we find that in Fig. \ref{fig:9} the median G of SPH, ETD, LTD, and IRR galaxies are 0.56, 0.50, 0.48, and 0.45, respectively. This follows a gradually decreasing trend, thus indicating that the luminosity distribution of the galaxy is gradually uneven with these categories. This is consistent with the conclusion of our previous works \citep{Zhou_2022,Fang_2023,Dai_2023,Song_2024}. The same trend had also been found by some other previous studies \citep{2008MNRAS.386..909C,2014ARA&A..52..291C}. Combined with the median trend of the $M_{20}$ parameter, the values of SPH, ETD, LTD, and IRR increase progressively. The upper and lower quartiles of the two parameters maintain the same trend, which indicates that all kinds of galaxies are well clustered together and galaxies of the same type have similar physical properties. They are distributed in a specific parameter space.

\begin{figure*}

        \includegraphics[width=1.5\columnwidth]{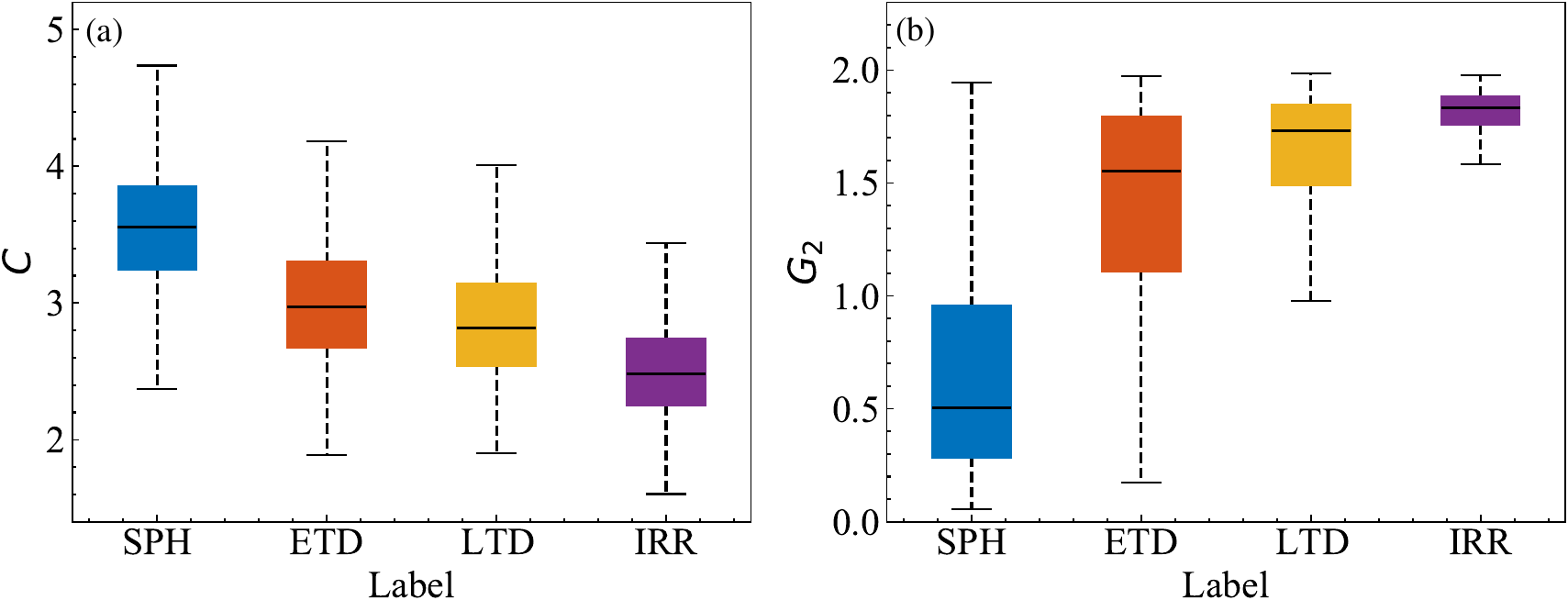}
 \centering
    \caption{Similar to Fig. \ref{fig:8} but for $C$ (left) and $G_{2}$ (right)
    It is evident that from SPHs to IRRs, the C of galaxies gradually decreases, while the $G_{2}$ gradually increases. Different types of galaxies have significantly different parameter distributions.}
    \label{fig:11}
\end{figure*}

Fig. \ref{fig:11} shows the distributions of different categories of galaxies in the space of parameters C and $G_{2}$. In panel (a), with the increase of parameter C, the median distribution of SPH, ETD, LTD, IRR, and galaxies gradually decreases to 3.55, 2.97, 2.82, and 2.48, respectively. In panel (b), the median values of $G_{2}$ from SPHs to IRRs are 0.51, 1.55, 1.73, and 1.83, respectively. With the increase of parameter $G_{2}$, the median of SPHs, ETDs, LTDs, and IRRs gradually increases, respectively. From Fig. \ref{fig:11}, we can analyze that from SPH to IRR, the compactness of the galaxy is decreasing. The more compact galaxies tend to be SPH or ETD, and the more sparse galaxies tend to be LTD and IRR.

\label{subsubsec:nonparametric}

\begin{figure*}
    \includegraphics[width=2\columnwidth]{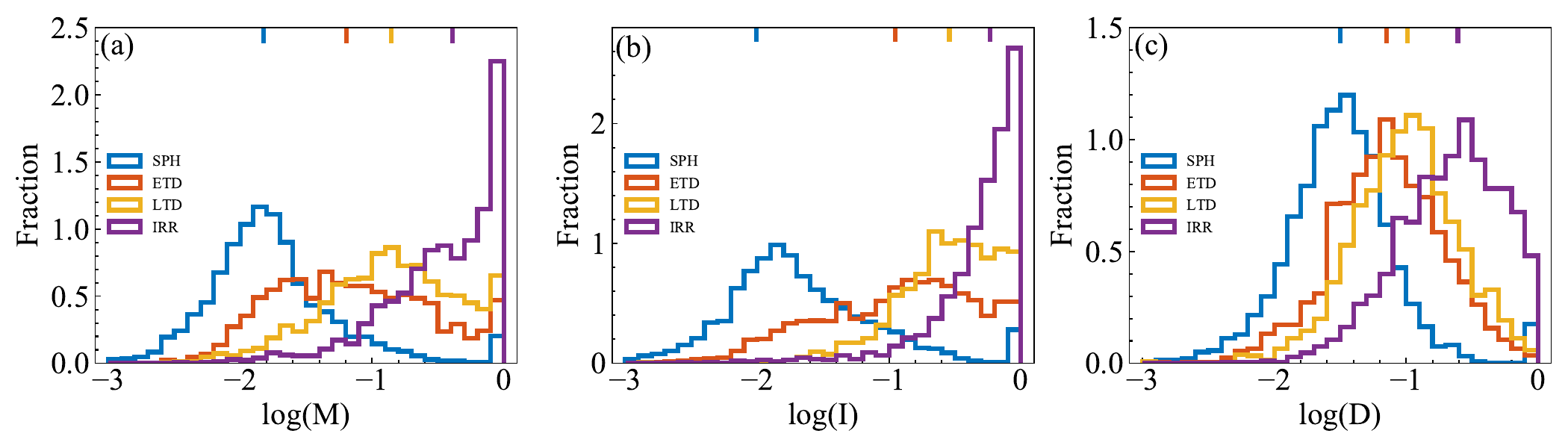}
    \caption{Distribution of M (left), I (middle), and D (right) for different types of massive galaxies. The bars at the top represent the median values of M, I, and D for different galaxy types. It is evident that from SPHs to IRRs, the I of galaxies gradually increases, and the D also gradually increases. There are significant differences in the M, I, and D parameters among different categories of galaxies, which indirectly proves the effectiveness of our classification. (SPH: blue, ETD: orange, LTD: yellow, IRR: purple)}
    \label{fig:12}
\end{figure*}

Fig. \ref{fig:12} suggests that as the ratio of light between two regions within galaxies strengthens and the distance between the image's light-weighted centre and the brightest peak increases, galaxies exhibit larger values of M, I, and D. Specifically, SPH galaxies exhibit smaller values of M, I and D. ETD, LTD, and IRR galaxies exhibit larger values of M, I and D. Different types of galaxies are clearly distinguished in parameter space, this validates the accuracy of our classification.

Compared to the original UML model \citep{Zhou_2022}, our enhanced UML model not only boasts superior classification behaviour, but also significantly reduces machine clustering results from 100 to 20 groups, thereby enhancing convenience for subsequent scientific research on specific galaxies. Notably, the classification between classes exhibits more distinct morphology, as illustrated in Fig. \ref{fig:5}. Moreover, qualitative analysis, as depicted in Fig. \ref{fig:7}, reveals that when the 1500-dimensional vector encoded by the larger model is projected onto a two-dimensional plane, approximately 80\% of the galaxy mappings are disconnected, underscoring sharper discrimination. Furthermore, our model has undergone parameter testing, including assessments based on M, I, D, C, and $G_{2}$, among others. These comprehensive tests validate the efficacy and advancement of our approach. Compared to traditional visual classification \citep{2023ApJ...946L..15K}, deep learning classification based on visual labels \citep{Dieleman_2015}, or datasets with available labels such as Galaxy Zoo 2 \citep{10.1093/mnras/stt1458}, the advantage of our classification model is that it does not rely on visual label data and does not require extensive training to achieve complete classification results. This reduces the bias caused by label data and greatly saves computational resources and time, providing a more efficient and convenient classification method for future scientific work. Lastly, our method of classifying into 20 categories is also beneficial for discovering new subcategories, which can facilitate subsequent scientific analyses.

\section{Summary} \label{sec:5}
In this paper, we propose an enhanced UML method that significantly expands the effective categories of machine classification from 100 classes to 20 classes. This improvement dramatically enhances the efficiency of visual classification and validates the applicability of our bagging-based multi-model voting strategy for large-scale data. The method can be summarized on the basis of three key aspects. (1) Preprocessing the source data using CAE and APCT techniques to reduce image noise and enhance rotation invariance. (2) Utilizing a pre-trained ConvNeXt model to encode the preprocessed image data, extract features, and further compress them through PCA dimensionality reduction. (3) Employing a bagging-based multi-model voting strategy, we successfully classified 53612 galaxies into 20 groups. Then, by visual inspection, we successfully classified 20 groups into 5 classes.

This enhancement elevates machine learning performance through visual classification by clustering galaxies into five highly similar feature-based classes: SPH, ETD, LTD, IRR, and UNC. Furthermore, significant differences exist among different galaxy types regarding these parameters. By utilizing visualization tools such as box plots and frequency distribution maps based on a parameter space analysis of classified galaxy morphology images, we can demonstrate their correspondence with various physical properties of galaxies while validating the reliability of our distribution.

Our improved UML method effectively clusters large-scale data using single-band images by grouping highly similar images into one class. This advancement enhances machine classification performance, while saving time in visual classification tasks and reducing labour consumption to a greater extent. Our future research will explore additional possibilities within machine learning methods. 

In the future, our endeavors in galaxy morphology classification will pivot around the efficient encoding of large-scale models. We will delve into various feature extraction methods, striving to continually enhance the accuracy of machine learning models. Additionally, we will venture into applying multimodal large-scale models in galaxy morphology classification. We will use dual coding contrastive learning technology to further improve feature extraction accuracy, while striving to further enhance the classification ability of the model and gradually reduce the number of machine clustering categories to 10, with an aim to improve the performance of scientific research. Additionally, we also look forward to using this technology to discover some interesting groups, such as galaxy pairs. Furthermore, we aim to streamline our algorithm into a more user-friendly and understandable software package, facilitating its utilization by future telescopes aboard the Chinese space station.

\begin{acknowledgements}
Financial support for this work is provided by the Strategic Priority Research Program of the Chinese Academy of Sciences (Grant No. XDB 41000000), the National Natural Science Foundation of China (NSFC, Grant No. 12233008, 11973038, 62106033, 12303017), the China Manned Space Project (No. CMS-CSST-2021-A07), the Cyrus Chun Ying Tang Foundations, and the Frontier Scientific Research Program of the Deep Space Exploration Laboratory. C.C.Z. acknowledges support from Yunnan Youth Basic Research Projects (202401AT070016; 20232904E030002), Z.S.L. acknowledges the support from Hong Kong Innovation and Technology Fund through the Research Talent Hub program (PiH/022/22GS). This work is also supported by Anhui Provincial Natural Science Foundation projects (2308085QA33).
\end{acknowledgements}
\bibliographystyle{aa}
\bibliography{ref}
\end{document}